\documentclass[
 reprint,
 superscriptaddress,
 amsmath,amssymb,
 aps,
 prl
]{revtex4-2}

\usepackage{graphicx}
\usepackage{dcolumn}
\usepackage{bm}
\usepackage[hidelinks,colorlinks=true,linkcolor=blue,citecolor=blue, anchorcolor = blue, urlcolor = blue]{hyperref}
\usepackage[mathlines]{lineno}

\usepackage{braket}
\usepackage{color}
\usepackage{float}

\newcommand{\affsiqse}{\affiliation{Shenzhen Institute for Quantum Science and Engineering, Southern University of Science and Technology, Shenzhen 518055, China}}
\newcommand{\affiqa}{\affiliation{International Quantum Academy, Shenzhen 518048, China}}
\newcommand{\affhfnl}{\affiliation{Hefei National Laboratory, Hefei 230088, China}}

\newcommand{\pauli}[1]{\sigma_{\mathrm{#1}}}
\newcommand{\yb}{^{171} \mathrm{Yb}^+}
\newcommand{\hg}{\mathrm{HG}_{01}}
\newcommand{\figref}[2]{Fig.~\ref{#1}\textcolor{blue}{#2}}

\newcommand{\exfigref}[2]{Extended Data Fig.~\ref{#1}\textcolor{blue}{#2}}

\begin{document}
\title{
    Scalable entangling gates on ion qubits via structured light addressing
    }

\author{Xueying Mai}
    \affsiqse
    \affiqa
\author{Liyun Zhang}
    \affiqa
\author{Qinyang Yu}
    \affsiqse
    \affiqa
\author{Junhua Zhang}
    \email{zhangjunhua@iqasz.cn}
    \affiqa
\author{Yao Lu}
    \email{luyao.phy@gmail.com}
    \affsiqse
    \affiqa
    \affhfnl

\date{\today}

\begin{abstract}
    A central challenge in developing practical quantum processors is maintaining low control complexity while scaling to large numbers of qubits. Trapped-ion systems excel in small-scale operations and support rapid qubit scaling via long-chain architectures. However, their performance in larger systems is hindered by spectral crowding in radial motional modes, a problem that forces reliance on intricate pulse-shaping techniques to maintain gate fidelities. Here, we overcome this challenge by developing a novel trapped-ion processor with an individual-addressing system that generates steerable Hermite-Gaussian beam arrays. The transverse gradient of these beams couples qubits selectively to sparse axial motional modes, enabling  to isolate a single mode as entanglement mediator. Leveraging this capability, we demonstrate addressable two-qubit entangling gates in chains up to six ions with fidelities consistently around 0.97, achieved without complex pulse shaping. Our method significantly reduces control overhead while preserving scalability, providing a crucial advance toward practical large-scale trapped-ion quantum computing.
\end{abstract}

\maketitle


The quest for large-scale quantum computation faces a critical challenge for current trapped-ion systems. While they hold records for gate fidelities and coherence times in small scales \cite{wang2017single, wang2021single, PhysRevLett.117.060504, PhysRevLett.117.060505}, both leading architectures in intermediate scale -- quantum-charge-coupled devices (QCCDs) and static large ion crystals \cite{kielpinski2002architecture, pino2021demonstration, PhysRevLett.97.050505, guo2024site, PhysRevX.15.021052} -- exhibit control complexity that grows prohibitively with qubit count \cite{bruzewicz2019trapped}. Such escalaing overhead threatens to outweigh their quantum advantages.
The most severe bottleneck arises in entanglement generation. The reliance on collective motional modes as entanglement mediators creates an intractable conflict: rapid scaling qubit counts requires compromising the motional isolation essential for high-fidelity operations \cite{PhysRevLett.117.060504, PhysRevLett.117.060505}. This trade-off currently limits the path toward practical large-scale trapped-ion processors.

The core challenge stems from motional mode crowding. In static ion crystals, increasing the ion count raises the density of motional modes, degrading gate fidelity due to uncontrolled residual couplings between qubits and spectator motional modes.
QCCD architectures address this by shuttling smaller ion chains, but require complicated trap designs with dense electrodes and precise potential control \cite{delaney2024scalable}. 
Alternatively, while large ion chains enable rapid qubit scaling, they demand increasingly complex pulse sequences to handle the crowded radial motional spectrum to preserving gate fidelity \cite{zhu2006arbitrary, PhysRevLett.112.190502, PhysRevLett.120.020501, PhysRevApplied.13.024022, blumel2021power, lu2019global, PhysRevLett.134.050602}, shifting the scaling challenge from hardware design to control system complexity.

While QCCD architectures inevitably face engineering challenges, long ion chains promise a breakthough by leveraging axial motional modes. The sparse axial mode spectrum enables isolation of single or few modes as entanglement mediators even in hundred-ion chains, significantly simplifying gate operations. However, individual-addressing laser systems necessary for long-chain processors must propagate perpendicular to the ion chain \cite{debnath2016demonstration,lu2019global, PRXQuantum.2.020343}, making them incompatible with axial motion couplings. Current implementations of axial-motion-mediated gates are restricted to global entangling operations \cite{PhysRevLett.106.130506, PRXQuantum.2.020343}, requiring resource-intensive shelving techniques to entangle arbitrary qubit pairs. Overcoming these limitation would provide a crucial pathway toward large-scale trapped-ion processors that maintain both high gate fidelity and low control complexity.

Emerging techniques in manipulating spatially structured light suggest potential solutions \cite{forbes2021structured}. The electric field profiles of structured light enable novel light-ion interactions \cite{verde2023trapped}. For instance, Laguerre-Gaussian beams carrying orbital angular momentum (OAM) can modify atomic selection rules when interacting trapped ions \cite{schmiegelow2016transfer, PhysRevLett.119.253203}. More remarkably, the transversal profile gradient of the electric field in such beams can couple to all motional degrees of freedom when an ion is positioned at the beam center \cite{PhysRevLett.129.263603}. These unique propoties establishes the first viable path toward simultaneously achieving scalable axial-mode coupling and individual qubit control.

\begin{figure*}[htbp]
    \centering
    \includegraphics{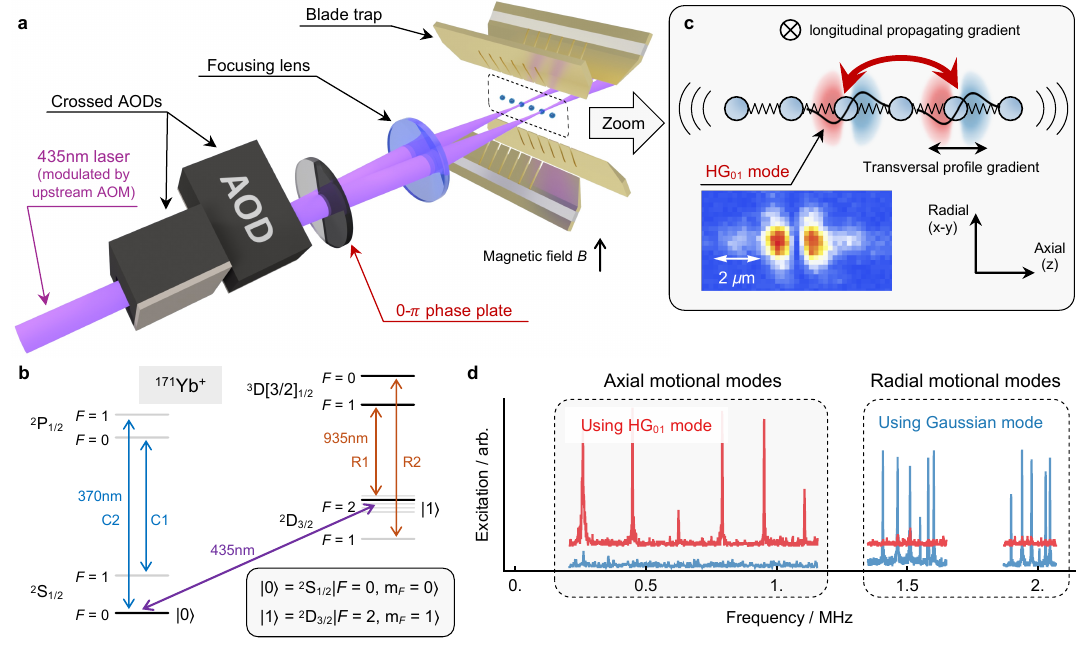}
    \caption{
        \label{fig:setup}
        \textbf{Trapped-ion quantum processor via Hermite-Gaussian-light addressing.}
        \textbf{a}. 
            Schematic of the experimental setup. A chain of $\yb$ ions is confined in a blade-type ion trap. Individual addressing beams, propagating perpendicular to the ion chain, are generated using orthogonally oriented AODs. A $0-\pi$ phase plate after the AODs converts the Gaussian beam into a $\hg$ mode at the focal plane.
        \textbf{b}. 
            Energy levels of single $\yb$ ion. Two levels from the $^2\mathrm{S}_{1/2}$ and $^2\mathrm{D}_{3/2}$ are used to encode an optical qubit. Transitions driven by 370~nm and 935~nm lasers are used for state preperation and measurement.
        \textbf{c}. 
            Spatial gradient of the $\hg$ mode. Conventionl laser beams exhibit longitudinal gradient along the propagating direction. Structured light fields such as the $\hg$ mode introduce an additional gradient arising from sharp spatial variation in field profile. By aligning the profile gradient along the axial direction of the ion chain, we can achieve qubit-qubit coupling mediated by axial motion (parallel to the chain axis). The inset shows the two-dimensional amplitude profile (excluding phase) of the focused $\hg$ mode as sampled by the ion.
        \textbf{d}. 
            Motional spectrum of a six-ion chain excited by Gaussian and $\hg$ mode beams. For the fundamental Gaussian mode, the propagating vector $\vec{k}$ projects onto both radial directions, exciting two set of radial motional modes in the spectrum. In contrast, the $\hg$ mode used in the experiments selectively excites only axial motional modes. The data of $\hg$ mode is vertically shifted to make it more visible.
        }
\end{figure*}

In this work, we develope a trapped-ion quantum processor incorporating an individual addressing system that utilizes spatially structured Hermite-Gaussian beams. This novel system enables both precise targeting of arbitrary ion qubits and selective coupling to axial collective motional modes, which is unattainable with conventional Gaussian beam addressing. We demonstrate efficient ground-state cooling and coherent manipulation of axial motion through this tailored light-motion coupling. 
Building on these capabilities, we implement entangling gates driven by Hermite-Gaussian beams in chains of up to six ions, achieving gate fidelities exceeding 0.97 without complex pulse modulation. Crucially, by utilizing both the field gradient and amplitude maxima of the structured light profile, our addressing system can realize a complete universal gate set. These advances establish a novel step toward scalable trapped-ion processors, significantly reducing the control complexity associated with the long-chain architecture.

\section{Results}
\subsection{Experimental setup}

we decipt our trapped-ion processor equipped with a Hermite-Gaussian individual addressing system in \figref{fig:setup}{a}. A chain of $\yb$ ions, with a tunable number of ions, is confined in a segmented blade trap. Unlike the conventional scheme that encodes the hyperfine qubit within the ground-state manifold of a single $\yb$ ion, we employ an optical qubit encoded in both the ground and metastable manifolds, defined as $\ket{0} = {^2\mathrm{S}_{1/2}}\ket{F = 0, m_F = 0}$ and $\ket{1} = {^2\mathrm{D}_{3/2}}\ket{F = 2, m_F = 1}$, with an energy gap $\omega_\mathrm{q}$ around 688~THz. This optical qubit is coherently manipulated using a narrow-linewidth 435~nm laser, as shown in \figref{fig:setup}{b}. A magnetic field of approximately 8~Gauss, aligned perpendicular the ion chain, is generated using permanent magnets. Since the energy gap of our qubit is first-order sensitive to the external magnetic field, a $\mu$-metal shield surrounding the vacuum chamber is employed to minimize external magnetic noise. All qubits can be initialized to the $\ket{0\dots0}$ state via standard optical pumping \cite{PhysRevA.76.052314}.
For a optical qubit, its state are measured via state-dependent fluorescence by driving the 370~nm C1 and C2 transitions along with the 935~nm R2 repump transition \cite{mai2024high}. Moreover, fluorescence from the ion chain is collected throught an objective lens with 0.4 numerical aperture (N.A.) and coupled into a multi-mode fiber array for site-resolved detection. The average detection fidelity for a single ion qubit is around 0.99 (see Methods for details).

To enable individual addressing of arbitrary ions in the chain, we employ a set of crossed acousto-optic deflectors (AODs) to generate and steer an array of laser beams without frequency shift \cite{kim2008doppler}. An additional acousto-optic modulator (AOM), placed upstream of the AODs (not shown in \figref{fig:setup}{a}), is used to rapidly modulate the addressing beams. In contrast to conventional addressing systems, here a spatial mode converter ($0-\pi$ phase plate) is placed right after the AODs, shaping the focused beam spot into an approximate first-order Hermite-Gaussian ($\mathrm{HG}_{01}$) mode. The focusing lens (N.A.$\sim$0.28) is placed on a three-dimensional motorized translation stage to fine tune the position of addressing beams. As illustrated in \figref{fig:setup}{c}, we characterize the profile of the focused $\mathrm{HG}_{01}$ beam using the ion itself, revealing a diameter of around $2~\mu$m (second-moment width, D4$\sigma$) in our setup  (see Methods for details).

Compared to previous demonstrations using Laguerre-Gaussian modes, the Hermite-Gaussian mode can be generated more easily with higher purity \cite{peshkov2023excitation}. Moreover, by aligning the dark slit of the $\hg$ mode perpendicular to the ion chain so that the transversal profile gradient is along the axial direction, we can achieve selective coupling to the axial motion of the ions. As shown in \figref{fig:setup}{d}, we can excite all the axial collective motional modes of a six-ion chain by using the $\mathrm{HG}_{01}$ beam, while conventionally only radial motion can be coupled by utilizing fundamental Gaussian mode.

\subsection{Manipulating single ion by laser in Hermite-Gaussian mode}

\begin{figure*}
    \centering
    \includegraphics{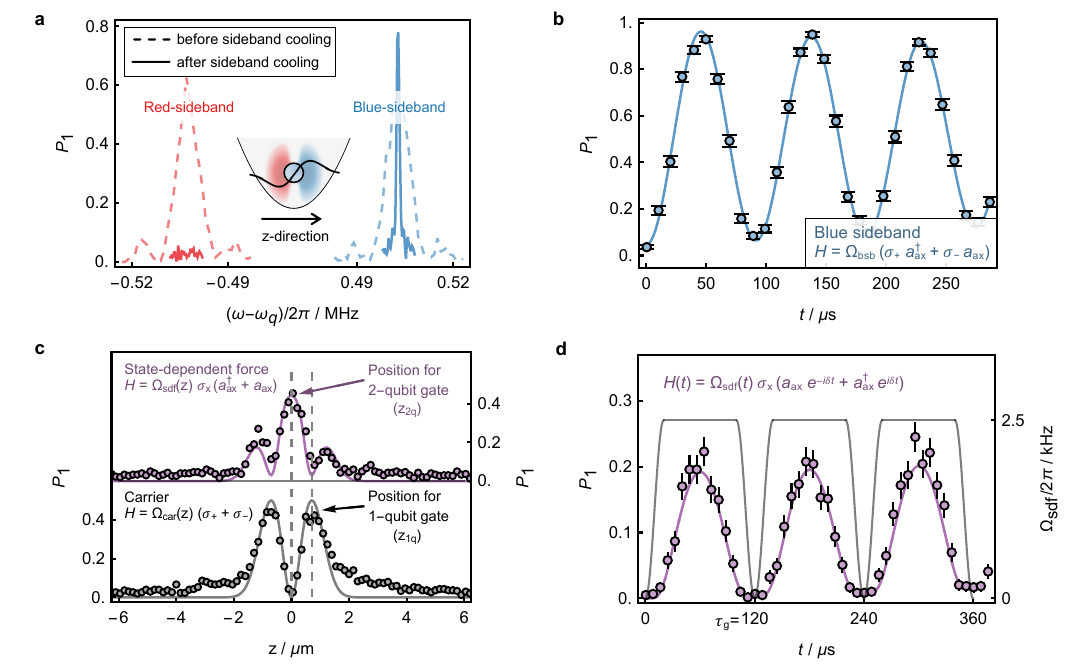}
    \caption{
        \label{fig:1q}
        \textbf{Coherent operations on single ion with laser in Hermite-Gaussian mode.}
        \textbf{a}.
            Sideband spectrum of axial motion after only Doppler cooling (dotted) and sideband cooling (solid). Sideband cooling suppresses the red sideband transition since the ion mostly occupies the $\ket{0}\ket{0}_\mathrm{ax}$ state ($\ket{n}_\mathrm{ax}$ denotes axial motional Fock states).
        \textbf{b}.
            Coherent blue-sideband oscillation after continuous sideband cooling. The Rabi flopping between $\ket{0}\ket{0}_\mathrm{ax}$ and $\ket{1}\ket{1}_\mathrm{ax}$ is observed. The fit (solid line) yields coupling strength of $\Omega_\mathrm{bsb} = 2\pi \times 5.45(1)$~kHz.
        \textbf{c}.
            Axial profile of $\hg$-mode laser field properties. After the sideband cooling, the ion is placed along the z-direction at different locations of the $\hg$ mode. The field gradient is measured via the state-dependent force under the situations of $\Omega_\mathrm{sdf}(z=z_\mathrm{2q}) = 2\pi \times 1.6$~kHz and the excitation duration $\tau_\mathrm{p} = 100~\mu$s. The field amplitude is probed through carrier transition by setting $\Omega_\mathrm{car}(z=z_\mathrm{1q}) = 2\pi \times 62.5$~kHz and $\tau_\mathrm{p} = 2~\mu$s. In our setup $z_\mathrm{1q} - z_\mathrm{2q} \approx 0.7~\mu$m. The solid lines show numerical simulations assuming a perfect $\hg$ mode profile.
        \textbf{d}.
            Qubit state oscillation under detuned state-dependent force. The frequency detuning and maxiaml coupling strength are set to be $\delta = 2\pi \times 10$~kHz and $\Omega_\mathrm{sdf}^\mathrm{(max)} = 2\pi \times 2.5$~kHz, respectively. Each 120~$\mu$s cycle incorporates 20~$\mu$s $\sin^2$-pulse shaping at both ends. The solid line represents numerical simulation including experimental imperfections.
        All error bars here and below represent 1$\sigma$ standard error, and most error bars are smaller than the marker size. 
    }
\end{figure*}

To investigate coherent manipulation of ion qubits using the $\hg$ mode, we begin with a single trapped $\yb$ ion, confined with an axial trap frequency of $\nu_\mathrm{ax} = 2\pi \times 0.502$~MHz. As shown in \figref{fig:1q}{a}, when the ion is placed at the dark slit of the $\hg$ mode, the axial-motion-related sideband transitions ($\omega_\mathrm{q} \pm \nu_\mathrm{ax}$) can be both excited after only Doppler cooling. To further cool the axial motion to its ground state, we implement continuous sideband cooling (CSC) via the red-sideband transition driven by the $\hg$-mode laser, assisted by the optical pumping process to pump the ion back to the $\ket{0}$ state. In \figref{fig:1q}{a}, we can find that the red-sideband transition is significantly suppressed after CSC, indicating near ground-state cooling of the axial motion. Under this situation, the Rabi oscillation of the qubit state under the blue sideband transition becomes visible, as illustrated in \figref{fig:1q}{b}. The average phonon number estimated from the oscillation is $0.02(2)$. 

To implement entangling operations between ion qubits, a state-dependent force is required in the Mølmer-Sørensen gate scheme \cite{PhysRevLett.82.1971, PhysRevLett.94.153602}. To examine this force with a single ion, we apply bi-chromatic frequency components, $\omega_\mathrm{q} \pm \mu$ (with $\mu \sim \nu_\mathrm{ax}$), to the addressing $\hg$-mode laser. It enables engineering the qubit-motion coupling of the form $ H =  \Omega_\mathrm{sdf} \sigma_\mathrm{x} (a_\mathrm{ax} e^{i\delta t} + a_\mathrm{ax}^\dagger e^{-i \delta t})$. 
Here, $\sigma_\mathrm{x,y,z}$ represent the Pauli matrices on single ion qubit, $a_\mathrm{ax}$ ($a_\mathrm{ax}^\dagger$) is the anihilation (creation) operator on the axial motional mode, $\Omega_\mathrm{sdf}$ is the strength of the qubit-motion coupling, and $\delta = \mu - \nu_{\mathrm{ax}}$ is the difference between the trap frequency and the beat-note frequency of the bi-chromatic fields.

Since the strength of the qubit-motion coupling is propotional to the field gradient of the addressing beam, it should be maximized when the ion is placed at the dark slit of the $\hg$ mode. We verify this by scanning the relative position between the ion and the addressing beam along the axial direction. In this test, the ion is initialized to the $\ket{0}$ state and driven by the resonant state-dependent force ($\delta = 0$). The resulting $\ket{1}$ state popultaion after excitation, $ P_{1} = \left[ 1 - \exp \left( {-2 \Omega_\mathrm{sdf}^2 \tau_\mathrm{p}^2} \right) \right]/2 $, dependent on the coupling strength $\Omega_\mathrm{sdf}$ and the excitation duration $\tau_\mathrm{p}$. By fixing $\tau_\mathrm{p}$, the spatial variant of the gradient can be profiled. As shown in \figref{fig:1q}{c}, the data clearly reveal that the gradient reaches its maximum at the dark slit.

By placing the ion at the dark slit and detuning the state-dependent force from resonance ($\delta \neq 0$), the qubit periodically decouples from its axial motion while acquiring an additional geometric phase. This phase can lead to entanglment between qubits when multiple ions are driven simultaneously. In \figref{fig:1q}{d}, we show the such periodic oscillation in the measured qubit state population at the experimental setting used in the subsequent entanglment generation.

Moreover, to implement a universal gate set for quantum computing, our addressing system leverages distinct features of the $\hg$ mode for both single- and two-qubit operations. The gradient maxima enable qubit-motion coupling for two-qubit gates, while the amplitude maxima is employed for single-qubit rotations.
As shown in \figref{fig:1q}{c}, the field amplitude of the addressing beam is characterized by driving the carrier transition. As expected, the excitation probability reaches a minimum at the dark slit, with two symmetric peaks appearing away from the center. Thus, in the following demonstrations, ions are aligned to the dark slit of each addressing beam for two-qubit entangling gates, and to one of the intensity peaks for single-qubit gates. We can also notice that the experimentally measured amplitude profile exhibits a long-tail distribution away from the center. This deviation arises from the intrinsic properties of the approximate $\hg$ mode generated using the $0-\pi$ phase plate, as discussed in the Methods section.

\subsection{Entangling gates in a three-ion chain}

\begin{figure*}
    \centering
    \includegraphics{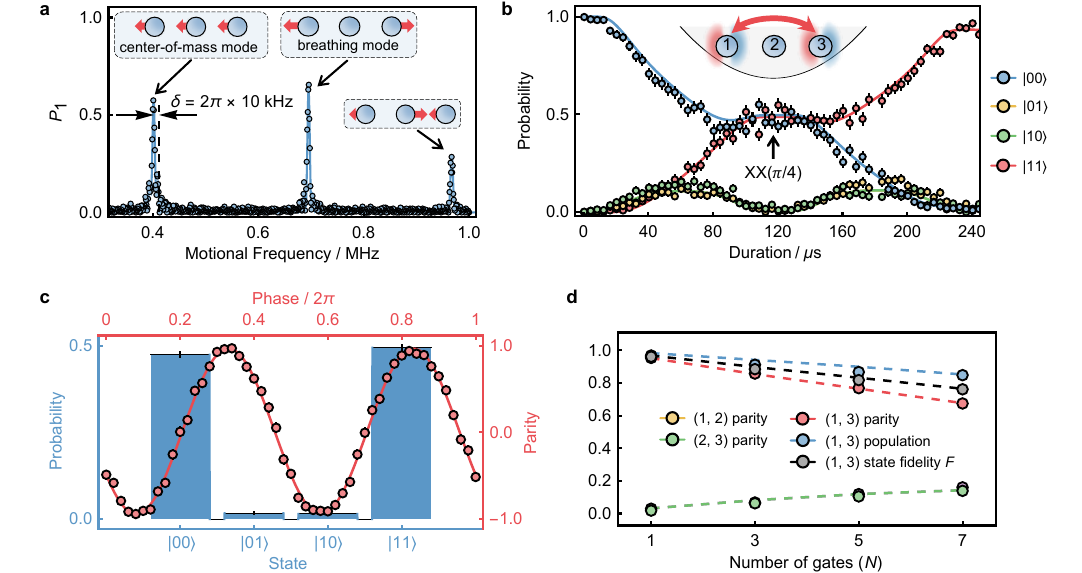}
    \caption{
        \label{fig:3q}
        \textbf{Characterization of two-qubit entangling gates in a three-ion chain.}
        \textbf{a}. 
            Axial motional spectrum of a three-ion chain, showing collective modes at $2\pi \times \{ 0.402, 0.696, 0.967 \}$~MHz. The beat-note frequency of the bichromatic fields is detuned by $\delta = 2\pi \times 10$~kHz from the COM mode. All open markers represent experimental data (addressing one outer ion), and the solid line corresponds to numerical simulations. Insets decipt the collective motion patterns.
        \textbf{b}.
            Qubit populations during two Mølmer-Sørensen gate cycles for the (1,3) ion pair. All qubits are initialized to $\ket{0}$ before operations. Laser power is optimized for $\phi=\pi/4$ at the gate duration of $\tau_\mathrm{g} = 120~\mu$s. Colored markers denote experimental results, while solid lines (in corresponding colors) represent numerical simulations incorporating experimental imperfections.
        \textbf{c}. 
            Bell-state fidelity measurement for the (1,3) ion pair. The blue histogram displays the state populations of the prepared entangled state, while the parity oscillation contrast quantifies off-diagonal coherence.
        \textbf{d}. 
            Bell-state fidelity vs. gate repetitions for the (1,3) ion pair. Colored markers indicate experimental results, and dotted lines show numerical simulations accounting for experimental error sources. Crosstalk-induced entanglement emerges in both (1,2) and (2,3) ion pairs, as evidenced by their increasing parity contrast with successive gate operations.
        }
\end{figure*}

After characterizing qubit-motion couplings in a single trapped ion, we extend our study to a three-ion chain for implementing entangling gates. The axial trap frequency is relaxed to $\nu_\mathrm{ax} = 2\pi \times 0.402$~MHz, yielding a nearest-ion spacing of 5.4~$\mu$m. As shown in \figref{fig:3q}{a}, three distinct collective motional modes are resolved. We employ the center-of-mass (COM) mode for mediating entangling gates due to its uniform coupling across all ion qubits. The beat-note frequency of the bi-chromatic fields is detuned by $2\pi \times 10$~kHz from the COM mode resonance, ensuring negligible coupling to other motional modes. Addressing and applying the state-dependent force to ($j$,$k$) ion pair enables entangling operations, $\mathrm{XX}_{j,k}(\phi) = \exp( - i \phi \pauli{x}^{(j)} \pauli{x}^{(k)})$, with a gate time of $\tau_\mathrm{g} = 120~\mu$s (including 20~$\mu$s $\sin^2$-ramp up/down) according to our settings. In \figref{fig:3q}{b}, we depict the state evolution of (1,3) ion pair under the state-state dependent force. A Bell state can be produced at the gate time by setting $\phi = \pi/4$ via laser power control.

We characterize the Bell state fidelity by measuring the population of the entangled state and the contrast of the parity oscillation \cite{sackett2000experimental}. For the (1,3) ion pair, as shown in \figref{fig:3q}{b}, the total population occupied in $\ket{00}$ and $\ket{11}$ states reaches 0.968(3), with a parity oscillation contrast of 0.953(6), corresponding to a fidelity of 0.960(3). 
We also generate Bell states for the (1,2) and (2,3) ion pairs, achieving fidelities of 0.949(4) and 0.953(3), respectively (see Supplementary Information for details).

To further benchmark gate performance, we measure the decay of the prepared Bell-state fidelity under repeated applications of an odd number of entangling gates. The results for the (1,3) ion pair are summarized in \figref{fig:3q}{d}. Note that, phase-varied $\pi/2$-rotations are applied to all three ions to extract parity oscillations of all ion pairs, revealing crosstalk-induced entanglement with the center ion. The measured error rate for the $\mathrm{XX}_{1,3}(\pi/4)$ operation is 0.035(1) per gate. The primary error source is dephasing of the 435~nm laser (coherence time around 3~ms), caused by residual frequency lock noise and fiber phase noise. Additional contributions include heating of the COM mode (around 250~phonon/s), and limited lifetime of the metastable $^2\mathrm{D}_{3/2}$ manifold (around 53~ms \cite{schacht2015171yb+}). The observed increase in parity contrast of (1,2) and (2,3) ion pairs reveals nearest-neighbor gradient crosstalk. Experimental measurements indicate around 1.5\% crosstalk, slightly exceeding the theoretical prediction of around 1\% (see Methods for details). A complete error budget is provided in Tab.~\ref{tab:error}. The numerically simulated gate error agrees well with the experimental result.

\begin{table}[htbp]
    \caption{
        \label{tab:error}
        Error budget for the entangling gate $\mathrm{XX}_{1,3}(\pi/4)$.
    }
    \begin{ruledtabular}
    \begin{tabular}{lcdr}
    \textrm{Error source} & \textrm{Simulated error}\\
    \colrule
    Laser dephasing &  $2.2\times 10^{-2}$ \\
    Heating of COM mode & $0.6\times 10^{-2}$ \\
    Gradient crosstalk &  $0.4\times 10^{-2}$ \\
    $^2\mathrm{D}_{3/2}$ lifetime &  $0.2\times 10^{-2}$ \\
    Pointing fluctuation &  $<1\times 10^{-3}$ \\
    Spectator modes &  $<1\times 10^{-6}$ \\
    \colrule
    Simulation sum &  $3.4 \times 10^{-2}$ \\
    Experimental error &  $3.5(1)\times 10^{-2}$ \\
\end{tabular}
\end{ruledtabular}
\end{table}

We further validate our error model by using the breathing mode to mediate the entangling gate for the (1,3) ion pair. In this configuration, gate errors arising from motional heating and gradient crosstalk are significantly suppressed. Experimental results demonstrate a gate error of 0.028(1), in agreement with numerical simulations predicting 0.027 (see Supplementary Information for details). 

\subsection{Extending to longer chains}

\begin{figure}
    \centering
    \includegraphics{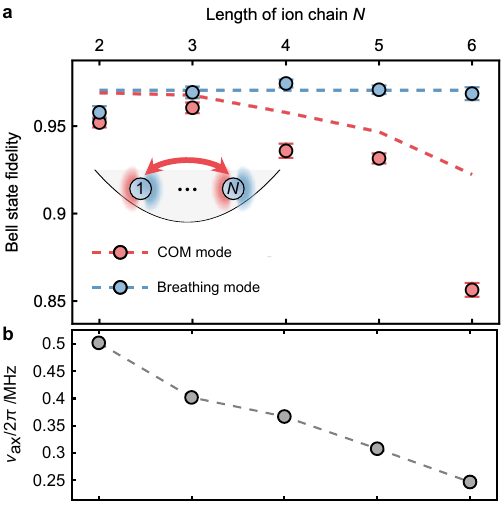}
    \caption{
        \label{fig:longchain}
        \textbf{Entanglement fidelities scaling with ion-chain length.}
        \textbf{a}. 
            Measured Bell-state fidelities for outermost ion pairs versus chain length, comparing COM mode (red) and breathing mode (blue) mediated gates. All colored markers represent experimental results, and dotted lines are obtain from numerical simulations including experimental imperfections. The Bell-state fidelities for COM-mode-mediated gates are 0.952(3), 0.960(3), 0.936(4), 0.932(3), 0.856(4) for two- to six-ion chains, respectively. While for breathing-mode-mediated gates, the Bell-state fidelities for two to six ions are 0.958(4), 0.969(3), 0.974(4), 0.971(3), 0.969(4), respectively.
        \textbf{b}.
            Axial trap frequencies for different chain lengths. For ion chains consisting of two to six ions, we set the axial trap frequencies to $2\pi\times \{0.502, 0.402, 0.367, 0.308, 0.247\}$~MHz. The uncertainties of all measured trap frequencies are less than 1~kHz.
    }
\end{figure}

To further assess the scalability of the $\hg$-driven entangling gates, we extended the system to up to six ions. As in the three-ion case, we employ the COM mode to mediate the entangling gates. Since all ions couple uniformly to the COM mode, we only benchmark gate performance on the outermost ion pair (1,$N$), where $N$ represents the length of ion chain. Such demonstrations also show the advantage of long-range connectivity in trapped-ion quantum processors.

As shown in \figref{fig:longchain}{a}, we measure the fidelity of prepared Bell states as the ion chain length increases. The fidelity remains above 0.95 for two- and three-ion chains, decreased to 0.93 for both four- and five-ion cases, and further dropped to 0.86 for the six ions. Meanwhile, as depicted in \figref{fig:longchain}{b}, the axial trap frequency is also lowered down from $2\pi \times 0.502$~MHz (two ions) to $2\pi \times 0.247$~MHz (six ions) to maintain nearly constant ion spacing. The observed fidelity degradation likely stems from increased motional heating, given the exponential scaling of heating rates with decreasing trap frequency.
Thus, we independently measure heating rates at different trap frequencies and include them into numerical simulations. The simulated fidelity trends show agreement with experimental data, except for a notable deviation in the six-ion case. We attribute this deviation to imperfect ground-state cooling due to the quite large heating rates (around 1600~phonon/s), which amplifies the heating-induce errors.

To further verify and mitigate heating impact, we implement entangling gates via the breathing mode, which exhibits a significantly lower heating rate (less than 10~phonons/s). In this case, we observe consistently Bell-state fidelities of $0.97$ for chains up to 6 ions. The residual errors are primarily attributed to laser dephasing noise and the finite lifetime of the $^2\mathrm{D}_{3/2}$ state, as discussed in the three-ion case.

Heating-induced gate errors present an inevitable challenge in trapped-ion quantum processors. Here we demonstrate one mitigation strategy by utilizing high-order motional modes as entanglement mediators. Alternative possible approaches include quantum control via multi-tone schemes to achieve heating-resilient gates \cite{valahu2022quantum, PhysRevA.111.042404}, as well as cryogenic systems to suppress heating rates by several orders of magnitude \cite{pagano2018cryogenic}.

We also find that the Bell state fidelity in the two-ion chain is slightly lower than expected with either mode as mediator. We attribute this extra error to off-resonant carrier coupling caused by the long-tail distribution of the amplitude profile in the $\hg$ mode produced by the phase plate. The same fidelity degradation occurs when entangling adjacent ion pairs in the three-ion chain, suggesting the same reason.

\section{Conclusion and Discussion}

In summary, we have demonstrated a novel individual addressing system that steers Hermite-Gaussian beam arrays to coherently manipulate and entangle ion qubits. By harnessing the transvesal profile gradient of the Hermite-Gaussian beam, we successfully employ the sparser axial motional modes to mediate entangling operations, eliminating the need for complex pulse modulation. This approach significantly reduces control complexity compared to conventional long-chain architectures. In chains up to six ions, we achieve the entangling gate fidelity exceeding 0.97, primarily limited by technical noise sources that can be systematically addressed. Our study provides a scalable pathway for trapped-ion quantum processors that enables efficient scaling to larger qubit numbers while minimizing control overhead.

While we demonstrate entangling gates using optical qubits with Hermite-Gaussian beams, this approach is directly applicable to hyperfine qubits. For hyperfine qubits, one Raman beam should be shaped into the $\hg$ mode, while the other remains in the fundamental Gaussian mode. Crucially, since the gradient for qubit-motion coupling originates from the spatial amplitude profile rather than beam propagation direction, both Raman beams can co-propagate. This configuration could significantly suppress optical path fluctuation noise.

In our current implementation, the approximate $\hg$ mode generated from $0-\pi$ phase plate inherently exhibits non-negligible crosstalk. For future improvement, alternative spatial mode-shaping techniques such as Spatial Light Modulators (SLMs), Digital Micromirror Devices (DMDs) \cite{shih2021reprogrammable, motlakunta2024preserving}, or laser-written waveguides \cite{sotirova2024low} could be explored to generate high-purity $\hg$ modes for low-crosstalk addressing.

The generation of high-purity $\hg$ modes also provides a new approach for breaking the entangling gate speed limit. Similar to standing-wave methods demonstrated in previous work \cite{PhysRevLett.131.220601, PhysRevLett.130.133201}, placing ions at the dark slit of $\hg$ beams can suppress the parasitic off-resonant carrier transitions in the Mølmer-Sørensen interaction, while maintaining the qubit-motion couplings. Furthermore, by simultaneously coupling multiple ion pairs to different axial motional modes, it is possible to implement parallel entangling gates with minimal control complexity \cite{figgatt2019parallel, zhu2023pairwise}. This scheme requires bi-chromatic fields with different beat-note frequencies to each addressing beams; thus it is particularly suitable for multi-channel-AOM-based addressing systems \cite{debnath2016demonstration, lu2019global}.

Future research could also explore interactions with other forms of structured light. For instance, second-order Hermite-Gaussian beams could enable purely dispersive qubit-motion coupling \cite{PRXQuantum.4.040342}, serving for single-shot phonon-resolved measurement in phonon networks \cite{chen2023scalable}. The flexibility in engineering diverse qubit-motion interactions may open new pathways for developing programmable hybrid quantum simulators incorporating both discrete and continuous variables \cite{PhysRevResearch.3.043072}. Such advances would further expand the toolbox of trapped-ion systems for practical quantum applications.

\bibliography{Refs}

\setcounter{figure}{0}
\renewcommand{\figurename}{Extended Data Fig.}

\section{Methods}

\
\subsection{Detection scheme for optical qubit}

\begin{figure}
    \centering
    \includegraphics{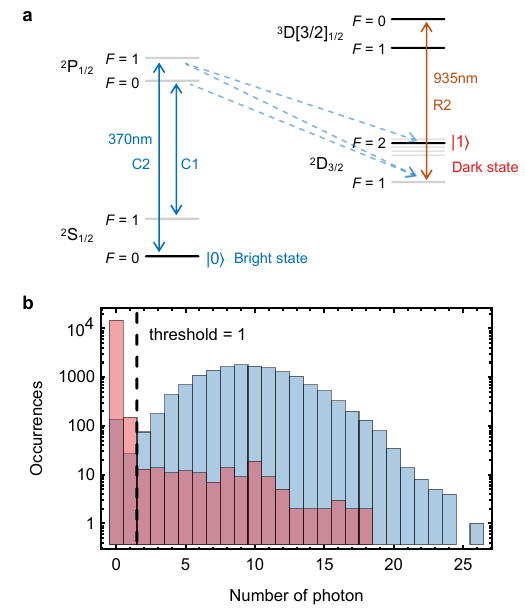}
    \caption{
        \label{fig:detection}
        \textbf{Detection scheme for $\yb$ optical qubit.}
        \textbf{a}. 
            Energy levels and transitions for detection process. We employ the 370 nm (C1 and C2) and 935 nm (R2) transitions for state-dependent fluorescence detection. The qubit state $\ket{0}$ (bright state) generates strong photon scattering, while $\ket{1}$ (dark state) remains non-fluorescent due to off-resonant to all detection lasers
        \textbf{b}.
            Photon count statistics. Histograms show photon counts from 15,000 experimental repetitions for states $\ket{0}$ (blue) and $\ket{1}$ (red). Setting a discrimination threshold at 1 photon yields bright and dark state fidelities of 0.9891(8) and 0.9909(8) respectively, resulting in a total detection fidelity of 0.9900(6).
        }
\end{figure}

The detection scheme for the $\yb$ optical qubit is a variant of the electronic shelving method. As shown in \exfigref{fig:detection}{a}, the optical qubit state $\ket{0} = {^2\mathrm{S}_{1/2}}\ket{F = 0}$ participates in a closed detection cycle as a bright state while the $\ket{1}$ state within the ${^2\mathrm{D}_{3/2}}\ket{F = 2}$ levels left dark, as previously demonstrated in \cite{mai2024high}. The main detection cycle is driven by the 370 nm C1 transition and enclosed by the 935 nm R2 transition. The 370 nm C2 transition is additionally driven to pump the $\ket{0}$ state population into the detection cycle, making it a bright state. In \exfigref{fig:detection}{b}, we illustrate photon statistics for both qubit states. The detection fidelity for a single ion is measured to be 0.9900(6). For multi-ion detection, we employ a multimode fiber array to simultaneously collect fluorescence across the ion chain. This configuration maintains the detection fidelity of 0.99 for up to six ions with negligible crosstalk.

Based on the above pre-calibrated detection errors, we apply post-processing error correction to remove detection errors \cite{shen2012correcting}. The raw detection results and detection error matrix are represented as $\mathbf{P}^\mathrm{meas} = \{ p^\mathrm{meas}_{0...0},...,p^\mathrm{meas}_{1...1} \}$ and $\mathbf{M}$, respectively. The true state populations $\mathbf{P}^\mathrm{real}$ are estimated by solving the least-squares minimization problem:
\begin{equation}
 \mathrm{min}~\| \mathbf{P}^\mathrm{meas} - \mathbf{M}.\mathbf{P}^\mathrm{real} \| _2,
\end{equation}
where $\|\cdot\|_2$ denotes the 2-norm. Given that the detection errors are consistent across all ion qubits and detection crosstalk is negligible, the $N$-qubit error matrix can be expressed as the tensor product of single-qubit error matrices $\mathbf{M}_\mathrm{1q}$,
\begin{equation}
    \mathbf{M} = \mathbf{M}_1\otimes...\otimes\mathbf{M}_N = \mathbf{M}_\mathrm{1q}^{\otimes N}.
\end{equation}

\subsection{Generation of Hermitian-Gaussian mode}
The ideal $\hg$ mode has a electric field distribution of
\begin{eqnarray}
    E(z) &=& E_0\times\mathrm{H}_1(\sqrt{2} \dfrac{z}{w_0}) \exp(- \dfrac{z^2}{w_0^2}) \nonumber\\
    &=& E_0\dfrac{2\sqrt{2}z}{w_0} \exp(- \dfrac{z^2}{w_0^2}),
\end{eqnarray}
where $\mathrm{H}_1(x)$ is the first-order Hermite polynomials, and $w_0$ is the beam radius defined by D4$\sigma$. The transversal profile gradient can be expressed as,
\begin{equation}
    \partial_z E(z) = E_0\dfrac{2\sqrt{2} (w_0^2 - 2z^2) }{w_0^3} \exp(- \dfrac{z^2}{w_0^2}).
\end{equation}

\begin{figure}
    \centering
    \includegraphics{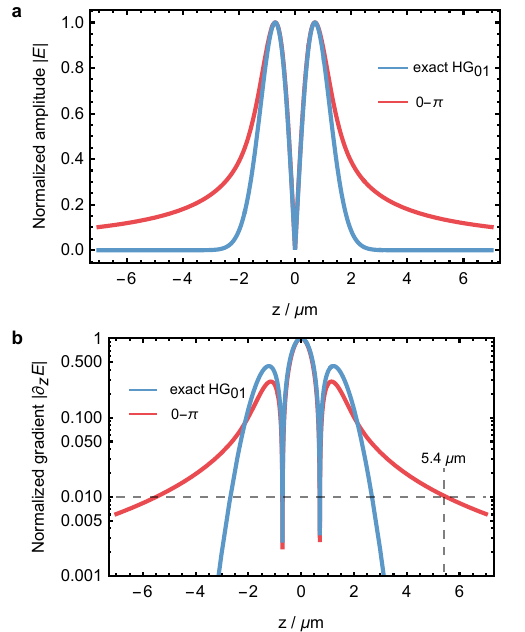}
    \caption{
        \label{fig:HG}
        \textbf{Comparison between exact $\hg$ mode and $0-\pi$ phase plate generated structured mode.}
        \textbf{a}. 
            Electric field amplitude profiles normalized to unity peak value. Normalized amplitude distribution $|E(z)|$ for both the $\hg$ and $0-\pi$ mode show double-peak features. The distance between two peaks is around 1.4~$\mu$m, corresponding to the experimental setting.
        \textbf{b}.
            Electric field gradient profiles, $|\partial_z E(z)|$, with maximal gradients normalized to unity. The $0-\pi$ mode exhibits around 1\% gradient crosstalk at 5.4~$\mu$m from the center.
        }
\end{figure}

In our demonstration, we employ a $0-\pi$ spatial mode convertor to transform the input Gassian mode into an approximate $\hg$ mode.
However, the generated structured mode (referred to as the $0-\pi$ mode), exhibits slight deviations from an ideal $\hg$ mode. As shown in \exfigref{fig:HG}{a}, we compare their electric field amplitude profiles along the z-direction. We can find that the $0-\pi$ mode displays a long-tail distribution compared to the $\hg$ mode, which is also experimentally confirmed, as shown in \figref{fig:1q}{c}.

This long-tail distribution in the amplitude profile also results in a corresponding long-tail in the gradient profile, as shown in \exfigref{fig:HG}{b}. In our three-ion chain configuration, the spacing of the nearest ion is around 5.4~$\mu$m. The intrinsic gradient crosstalk of the $0-\pi$ mode is around 1\%, while our expeirmental meansurments yield around 1.5\%.

\textbf{Data availability}: 
  The data that support the findings of this study are available from the authors upon request.
\

\textbf{Acknowledgements}:
    We gratefully thank Dapeng Yu for providing critical support to this project.
    We appreciete Heng Shen for carefully reading our manuscirpt and providing critical comments.
    This work was supported by the National Science Foundation of China (Grants No.~12004165 and 12104212), the Guangdong Basic and Applied Basic Research Foundation (Grant No. 2022B1515120021) and the Shenzhen Science and Technology Program (Grants No.~RCYX20221008092901006).
\

\textbf{Author contributions}:
    X.-Y.M., L.-Y.Z., Q.-Y.Y., J.-H.Z. and Y.L. developed the experimental system. 
    J.-H.Z. developed the RTMQ framework for the experimental control system.
    X.-Y.M. and Y.L. designed the experiments, took and analysed the experimental data.
    Y.L. and J.-H.Z. conceived the project. 
    Y.L. supervised the project.
    All authors contributed to write the manuscript.
\
  
\textbf{Competing interests}:
    The authors declare that there are no competing interests.
\

\textbf{Author information}:
    Correspondence and requests for materials should be addressed to J.-H.Z. and Y.L.

\setcounter{equation}{0} 
\setcounter{figure}{0}
\renewcommand{\figurename}{Fig.}
\setcounter{table}{0} 
\renewcommand{\theparagraph}{\bf}
\renewcommand{\thefigure}{S\arabic{figure}}
\renewcommand{\thetable}{S\arabic{table}}
\renewcommand{\theequation}{S\arabic{equation}}
    
\onecolumngrid
\flushbottom
\newpage
    
\section*{Supplementary Information}

\section{Entanglment in three-ion chain}

\begin{figure*}[htbp]
    \centering
    \includegraphics{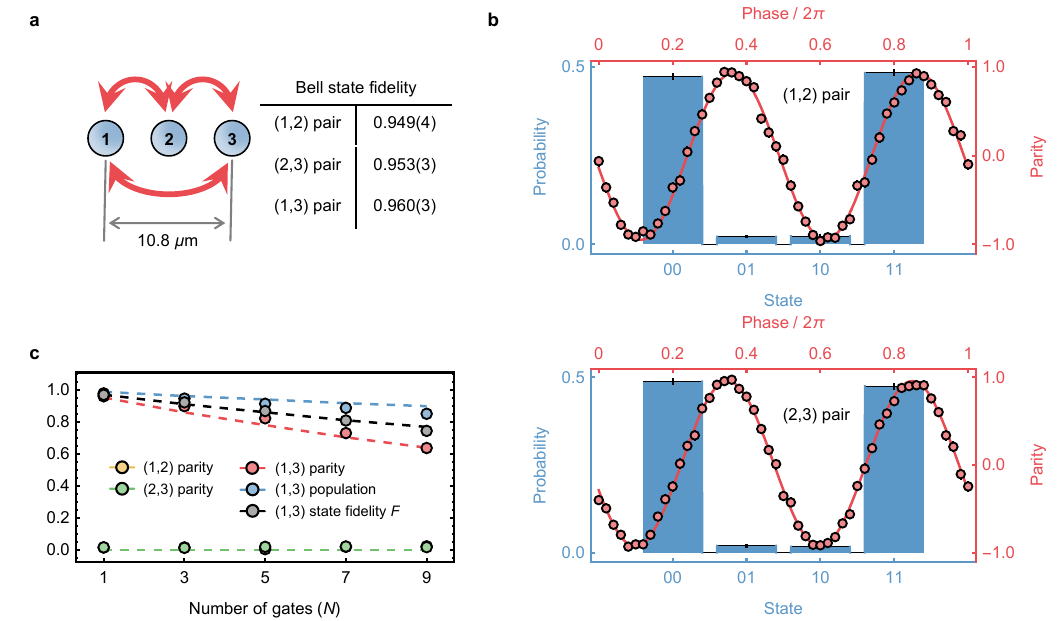}
    \caption{
        \label{fig:3qdata}
        \textbf{Entanglement generation in the three-ion chain.}
        \textbf{a}. Measured Bell-state fidelities for all ion pairs.
        \textbf{b}. Bell-state characterization for (1,2) (top) and (2,3) (bottom) pairs. The blue histogram displays the state populations, while the parity oscillation contrast quantifies off-diagonal coherence.
        \textbf{c}. Breathing-mode-mediated Bell-state fidelity versus gate repetitions for the (1,3) pair. Colored markers indicate experimental results, and dotted lines show numerical simulations accounting for experimental error sources.
        }
\end{figure*}

In the maintext we presents the detailed data of Bell-state fidelity for the (1,3) ion pair, \figref{fig:3qdata}{a-b} shows supplementary data for nearest-neighbor (1,2) and (2,3) pairs.
The measured fidelities for these adjacent pairs are systematically lower than those of the outermost (1,3) pair. As mentioned in the maintext, we attribute the fidelity difference to the long-tail induced field amplitude crosstalk for structured beams generated by $0-\pi$ phase plate. This amplitude crosstalk induces the off-resonant carrier coupling that degrade gate performance, which is also present in the two-ion chain case.

In \figref{fig:3qdata}{c}, we benchmark the $\mathrm{XX}_{1,3}(\pi/4)$ gate mediated by the breathing mode, implemented with a detuning of $2\pi\times 1/120$~kHz from the breathing mode frequency due to limited 435 nm laser power. Thus the gate time increases to $140~\mu$s (including 20~$\mu$s $\sin^2$-ramping up/down). The breathing mode’s inhomogeneous motional pattern suppresses heating effects and naturally decouples the center ion, significantly reducing gate errors from both motional heating and gradient crosstalk compared to the COM-mode-mediated case.
A complete error budget is provided in Tab.~\ref{tab:error2} 

\begin{table}[htbp]
    \caption{
        \label{tab:error2}
        Error budget for the entangling gate $\mathrm{XX}_{1,3}(\pi/4)$ mediated by breathing mode.
    }
    \begin{ruledtabular}
    \begin{tabular}{lcdr}
    \textrm{Error source} & \textrm{Simulated error}\\
    \colrule
    Laser dephasing &  $2.46\times 10^{-2}$ \\
    $^2\mathrm{D}_{3/2}$ lifetime &  $0.24\times 10^{-2}$ \\
    Pointing fluctuation &  $<1\times 10^{-3}$ \\
    Spectator modes &  $<1\times 10^{-6}$ \\
    \colrule
    Simulation sum &  $2.7 \times 10^{-2}$ \\
    Experimental error &  $2.8(1)\times 10^{-2}$ \\
\end{tabular}
\end{ruledtabular}
\end{table}

\end{document}